\documentclass[runningheads,a4paper,oribibl]{llncs}

\usepackage[british]{babel}
\usepackage[utf8]{inputenc} % ignored in xelatex
\usepackage[T1]{fontenc}
\usepackage{microtype}

\usepackage{url}
\makeatletter
\g@addto@macro{\UrlBreaks}{\UrlOrds}
\makeatother
\usepackage{xcolor}

\usepackage[
bookmarks=false,
breaklinks=true,
colorlinks=true,
linkcolor=black,
citecolor=black,
urlcolor=black,
pdfpagelayout=SinglePage,
pdfstartview=Fit
]{hyperref}
\usepackage[all]{hypcap}

\usepackage[%
  backend=biber,
  style=numeric,sortcites,sorting=none,
]{biblatex}
\usepackage{amsmath,amssymb}
\usepackage{enumitem}
\setlist[description]{style=standard}
\usepackage{calc}

\usepackage{tikz}
\usepackage{pgfplots}
\usepackage{pgfplotstable}
\usepgfplotslibrary{statistics}
\pgfplotsset{%
  compat=1.12,
}

\usepackage{tabularx}
\usepackage{graphicx}
\usepackage{subcaption}
\def\tightlist{}
\usepackage{longtable}
\usepackage{booktabs}

\usepackage{siunitx}

\usepackage{csquotes}

\newcommand{\landau}[1]{\ensuremath{\mathcal{O}\left(#1\right)}}

\DeclareSourcemap{
 \maps[datatype=bibtex]{
   \map{
     \step[fieldsource=doi,
       match=\regexp{\{\\_\}},
       replace={2}]
     \step[fieldsource=url,
       match=\regexp{\{\\_\}},
       replace={_}]
   }
 }
}

\DeclareTextCommandDefault{\nobreakspace}{\leavevmode\nobreak\ }
\newcommand\cS{{\mathbb{S}}}

\newcommand\cSs{{\widehat{\mathbb{S}}}}

\usetikzlibrary{backgrounds,calc,fit,decorations.pathreplacing}
\usetikzlibrary{positioning,shapes,shadows,arrows,decorations.pathmorphing}

\begin{document}

\title{Secure and Trustable Distributed Aggregation based on Kademlia}
%If Title is too long, use \titlerunning
%\titlerunning{Short Title}

%Single insitute
\author{Stéphane Grumbach \and Robert Riemann}
%If there are too many authors, use \authorrunning
%\authorrunning{First Author et al.}
\institute{Inria Grenoble Rhône-Alpes
\newline \email{\{stephane.grumbach,robert.riemann\}@inria.fr}}

%Multiple insitutes
%Currently disabled
%
\iffalse
%Multiple institutes are typeset as follows:
\author{Firstname Lastname\inst{1} \and Firstname Lastname\inst{2} }
%If there are too many authors, use \authorrunning
%\authorrunning{First Author et al.}

\institute{
Insitute 1\\
\email{...}\and
Insitute 2\\
\email{...}
}
\fi
                        
\maketitle

\begin{abstract}
Aggregation of values that need to be kept confidential while
guaranteeing the robustness of the process and the correctness of the
result is required in an increasing number of applications. We propose
an aggregation algorithm, which supports a large spectrum of potential
applications including complex voting protocols. It relies on the
distributed hash table Kademlia, used in BitTorrent, for pseudonymous
communication between randomly predetermined peers to ensure a high
degree of confidentiality which does not solely relies on cryptography.
The distribution of data and computation limits the potential for data
breaches, and reduces the need for institutional trust. Experimental
results confirm the complexity of \(\landau{\log n}\) for \(n\) peers
allowing for large-scale applications.
\end{abstract}


\section{Introduction}\label{sec:introduction}

An increasing number of applications require aggregation of values that
should not be revealed, for various aspects of privacy protection. They
include personalized services related to domotic, smart cities, or
mobility for instance that are blooming today, while revealing security
breaches. Confidentiality protecting aggregation is of even greater
importance for online voting. We demonstrate that peer-to-peer systems
offer great promises for such aggregations, because they limit the
potential for data breaches and simplify the essential question of
trust.

This paper presents \textsc{Advokat}, a distributed protocol for
confidential aggregation of inputs produced by large sets of peers. It
relies on the distributed hash table Kademlia \autocite{maymounkov2002},
that offers both an overlay network to organize peers, as well as a tree
structure to compute the aggregation. Kademlia is a robust and scalable
technique which is used in particular by BitTorrent
\autocite{cohen2008}. The proposed protocol integrates also techniques
from Bitcoin \autocite{bitcoin08} and BitBallot \autocite{reimert2016}.

Voting is the main privacy preserving aggregation realized with
pre-digital technologies. Paper-based voting protocols offer an
unmatched solution to satisfy often contradicting though essential
properties, such as secrecy of the ballot, correctness of the tally and
verifiability. Moreover, the possibility given to voters to participate
in the supervision on-site of both the casting and tallying procedures
ensures trust. No expert knowledge is required to understand the
protocol and its verification procedure. Thus, no trust in organizing
authorities is necessary. Paper-based voting owes its robustness to its
independence from institutional trust. Our objective is to transfer as
much as possible these properties in the online world, while offering
new properties not available in the classical setting, such as remote
participation as well as the capacity to launch a new aggregation.

The case of voting protocols is particularly interesting due to its
conflicting, but essential security requirements. On one hand, the
eligibility of every voter to cast a ballot must be ensured, while, on
the other hand, no link can be established between a given ballot and
the corresponding voter. Furthermore, the final tally must be
verifiable. Distributed protocols are promising for voting since they
allow to reduce the reliance on trust and open new prospects for
verification. The various tasks are carried out collectively in a
peer-to-peer manner by the participants, much like voters in paper-based
voting.

We assume the existence of an administrator trusted to certify the
eligibility of peers. Supported by a tracker, eligible peers join a
Kademlia DHT that provides a tree-like overlay network in which peers
are assigned to random leaves. Peers pull inputs, and later input
aggregates, from other peers close to them in the tree overlay, which
allows to compute aggregates for all ancestor nodes up to the root. The
strategy resides on pulling versus pushing for dissemination.
Cryptographic signatures are used to authorize peers to pull in
different subtrees.

Although several protocols propose a distributed aggregation over
tree-like overlay networks
\autocites{zhang2003}{vanrenesse2004}{cappos2008}{artigas2006a}, to the
best of our knowledge, the proposed algorithm is the first to consider
eligibility, confidentiality, scalability and verifiability at once. The
DASIS protocol \autocite{albrecht2004} balances the Kademlia tree by
routing joining peers to less populated subtrees. Internally, the
subtree size is computed in a similar fashion to our approach, but no
security measures are introduced. A distributed, Kademlia-based voting
protocol to rank the quality of BitTorrent content has been proposed
\autocite{evseenko2014}. However, confidentiality and eligibility are
not addressed.

Using distributed protocols for voting is a very natural idea to avoid
concentration of power. Common building blocks, like blind signature
schemes \autocite{chaum1983}, Mix Networks \autocite{chaum1981} or
threshold decryption \autocite{gennaro1999} exercise decentralization on
a small scale. Many classical online voting protocols employ already a
set of authorities \autocites{benaloh1986}{fujioka1993}{ibrahim2003} to
achieve privacy. However, they assume trust in the authorities and the
aggregation is generally centralized, rendering the protocols vulnerable
to DDoS attacks and data breaches of global impact for instance.

Various
efforts\footnote{\raggedright Blockchain-based voting techniques include:\
\url{http://votem.com},
\url{http://cryptovoter.com},
\url{http://votosocial.github.io},
\url{http://followmyvote.com},
\url{http://bitcongress.org},
\url{http://github.com/domschiener/publicvotes}
} are ongoing to propose distributed online voting protocols, based on
the Bitcoin blockchain \autocite{bitcoin08}, that does not require
trusted authorities. Still, published results are sparse.
\autocite{zhao2015} describes a protocol for a binary majority voting to
determine the receiver of a voter sponsored Bitcoin payment.

The \emph{SPP} protocol \autocite{gambs2011}, based on Secure
Multi-Party Computation, partitions the aggregation over a tree
hierarchy of peers of which a random set of peers serves as authorities
to carry out the final decryption step. \emph{DPol}
\autocite{guerraoui2012} and its extension \emph{EPol}
\autocite{hoang2014} are similar to our protocol in that the aggregation
is distributed to all peers and for their renunciation of cryptography.
However, their message complexity does not allow for large-scale
elections.

The aggregation protocol is evaluated with respect to the security
properties used for centralized voting protocols such as FOO
\autocite{fujioka1993}, and to scalability properties used for
distributed aggregation protocols such as SPP \autocite{guerraoui2012}.
We consider eligibility, confidentiality (secrecy), completeness and
correctness, verifiability, and complexity in terms of messages, memory
and time.

The paper is organized as follows. In the next section, we present the
general setting of the protocol. The basic aggregation is shown in
Section~\ref{sec:basic-aggregation}, while the recursive process that
takes advantage of the tree overlay of the Kademlia DHT is shown in
Section~\ref{sec:recursive-aggregation}. Then, in
Section~\ref{sec:robust-aggregation}, the recursive aggregation process
is extended to allow a minority of dishonest Byzantine peers. Several
desirable security and complexity properties are sketched in
Section~\ref{sec:properties}. The provided confidentiality is
experimentally examined in Section~\ref{sec:simulation} by means of a
simulation.

\section{Aggregation Protocol}\label{sec:aggregation}

The protocol relies on \emph{peers}, an \emph{administrator} and a
\emph{tracker}. The administrator is entrusted to certify the
eligibility of peers. For this purpose, we assume an authenticated,
tamper-resistant communication channel between the administrator and
each peer, e.g.~using an existing public key infrastructure.

Once certified, \emph{peers} join a distributed hash table (DHT) that is
mainly used to find other peers, but allows also to retrieve and store
data. We choose the Kademlia DHT \autocite{maymounkov2002} whose
tree-like network overlay is well-suited for aggregations. Like in
BitTorrent, a \emph{tracker} is employed to provide an initial peer as
an entry point. Peers communicate via pairwise channels assumed to be
confidential and authenticated to the degree of a peer pseudonym, e.g.~a
public IP address.

We use the following notations adapted from \autocite{fujioka1993}:

\begin{longtable}[]{@{}rl@{}}
\toprule
\(A\) & Administrator\tabularnewline
\(P_i\) & Peer, \(i\)-th out of \(n\)\tabularnewline
\(a_i\) & Initial aggregate of peer \(P_i\)\tabularnewline
\((pk_i,sk_i)\) & public and private key pair of peer
\(P_i\)\tabularnewline
\(\eta(m)\) & Hashing technique for message \(m\),
e.g.~SHA-1\tabularnewline
\(\sigma_i(m)\) & Peer \(P_i\)'s signature scheme using
\((pk_i,sk_i)\)\tabularnewline
\(\sigma_A(m)\) & Administrator's signature scheme\tabularnewline
\(\chi(m,r)\) & Blinding technique for message \(m\) and random number
\(r\)\tabularnewline
\(\delta(s,r)\) & Retrieving technique of blind signature\tabularnewline
\bottomrule
\end{longtable}

\noindent
The proposed protocol follows the following structural steps:

\begin{description}
\tightlist
\item[Preparation]
Peers create personal public and private key pairs and send
authorization requests with their blinded public key to the
administrator.
\item[Administration]
Once for each peer, the administrator signs the peer's blinded public
key without learning it and sends the signature to the peer.
\item[Aggregation]
Supported by the tracker, peers join the tree-like overlay network of
Kademlia. Then, peers assign their \emph{initial aggregate} to their
leaf node and compute collectively the \emph{root aggregate} from all
initial aggregates using the \emph{distributed aggregation algorithm}.
This requires the computation of \emph{intermediate aggregates} for all
their ancestor nodes in the Kademlia tree.
\item[Evaluation]
On fulfilment of a well-defined verification criteria, peers accept
their root aggregate as \emph{final root aggregate}. The outcome
(e.g.~election result) is eventually derived from the final root
aggregate.
\end{description}

\noindent  In the preparation step, each peer \(P_i\) generates on it's
own authority a public and private key pair \((pk_i,sk_i)\) to sign
messages with \(\sigma_i(m)\). To limit the number of valid keys to one
per eligible peer, the public key must have the signature of the
administrator \(A\) \autocite{gambs2011}. As in FOO
\autocite{fujioka1993}, a blind signature scheme \autocite{chaum1983} is
used to ensure that \(A\) cannot recognize peers after the
administration step. \(P_i\) randomly chooses a blinding factor \(r_i\),
computes its blinded public key \(b_i = \chi(pk_i,r_i)\) and sends it to
\(A\) using the authenticated, tamper-resistant channel.

In the administration step, \(A\) ensures to sign only a unique \(b_i\)
for each \(P_i\) and responds to \(P_i\) with its signature
\(s_i = \sigma_A(b_i)\). Eventually, \(P_i\) can retrieve the
\emph{authorization token} \(t_i = \delta(s_i,r_i)\). \(A\) does not
intervene any further once all eligible peers have acquired their
authorization or a time-out has elapsed.

During the aggregation step, all peers run the distributed aggregation
algorithm, that is presented hereafter in
Section~\ref{sec:basic-aggregation} and \ref{sec:recursive-aggregation}.

\section{Basic Aggregation}\label{sec:basic-aggregation}

The aggregation algorithm allows to implement various kinds of
confidential aggregations. In particular, with standard security
requirements slightly weakened, it supports a large spectrum of voting
systems.

\emph{Aggregates} are values to be aggregated, whether initial
aggregates, constituting inputs from peers, or intermediate aggregates
obtained during the computation. The specification of the aggregation
algebra is formulated below. We then introduce the \emph{aggregate
container} allowing to attach meta-information to aggregates that is
used to position them in the tree and ensure verifiability.

We introduce an algebra whose operation applies to \emph{aggregates},
which are aggregated during the computation of the operation. In the
case of a vote, aggregates correspond to ballot boxes filled with
ballots, and the operation is the union of sets. The data structure can
be adapted to different applications with different aggregation
functions, such as average, majority voting, etc.

We consider a set \(\mathbb{A}\) of aggregates. The aggregation
operation, \(\oplus\), combines two \emph{child aggregates} to a
\emph{parent aggregate} in \(\mathbb{A}\). \emph{Initial aggregates},
corresponding to peer inputs, are not computed, but provided by the
peers. We assume that the
operation~\(\oplus: \mathbb{A}\times\mathbb{A} \mapsto \mathbb{A}\) is
commutative and associative.

Consider for illustrative purposes the algebra for the \emph{Plurality
Voting} (PV). Peers, or here more precisely voters, choose one out of
\(d\) options, that are modeled in the algebra with initial aggregate
vectors \((e_1,\ldots,e_d)\) in \(\mathbb{A} = \mathbb{N}^d\), with
\(\sum_{x=1}^d e_x = 1\). The operation \(\oplus\) is simply vector
addition in \(\mathbb{A}\). The root aggregate
\(a_R = (n_1,\ldots,n_d)\) with \(\sum_{x=1}^d n_x = n\) indicates how
many peers \(n_x\) have chosen each option. The option \(x\) with the
highest \(n_x\), hence plurality, corresponds to the vote outcome. The
system can be easily extended to \(\mathbb{A} = \mathbb{Q}_+^d\) to
support vote splitting between two or more options. The Manhattan norm
is used to ensure the validity of initial aggregates \(a_i\) with
constant weight: \(\|a_i\|\).

More complex voting systems such as for instance the \emph{Alternative
Voting} and the \emph{Single Transferable Voting} systems can easily be
encoded. In both cases, voters have to rank options. Every ranking of
\(d!\) possible rankings in total can be interpreted as one option in
the PV algebra. The set of aggregates \(\mathbb{A}\) consists of vectors
\(\mathbb{A} = \mathbb{N}_0^{d!}\) and the operation is again vector
addition. Note that alternative, more compact encodings can be defined
for efficiency reasons.

The aggregation algorithm relies on meta-information of an aggregate
\(a\) that is in general not directly involved in the aggregate
computation, and constitutes together with \(a\) the aggregate container
of \(a\):

\begin{longtable}[]{@{}rl@{}}
\toprule
\(h\) & hash \(\eta(\cdot)\) of the aggregate container without
\(h\)\tabularnewline
\(a\) & aggregate\tabularnewline
\(c\) & counter of initial aggregates in \(a\),
\(c = c_1 + c_2\)\tabularnewline
\(c_1\), \(c_2\) & counter of initial aggregates of child
aggregates\tabularnewline
\(h_1\), \(h_2\) & container hashes of child aggregates\tabularnewline
\(\cSs(x,d)\) & identifier of subtree whose initial aggregates are
aggregated in \(a\)\tabularnewline
\bottomrule
\end{longtable}

\noindent  The counter \(c\) allows to detect protocol deviations and to
measure the number of initial aggregates in the root aggregate that can
be compared to \(n\) \autocite{vanrenesse2004}.

The aggregate container hash \(h\) depends on its child aggregate hashes
\(h_1,h_2\). As such, a chain of signatures is spanned reaching from the
root or any intermediate aggregate down to the initial aggregates of the
peers. Also employed in the Bitcoin blockchain, this technique ensures
that the sequence of aggregate containers is immutable.

\section{Recursive Aggregation over the Kademlia Binary
Tree}\label{sec:recursive-aggregation}

The aggregation protocol relies on the Kademlia DHT that establishes a
binary tree overlay network in which each peer \(P_i\) is assigned to a
leaf node. Using the aggregation operator \(\oplus\), peers compute the
intermediate aggregate for all the parent nodes from their corresponding
leaf up to the root node of the tree. The aggregates used to compute any
intermediate aggregate of a given tree node are those of its child
nodes. Hence, aggregates have to be exchanged between peers of
\emph{sibling subtrees}, i.e.~subtrees whose roots have the same parent.
Kademlia is not used solely to discover other peers, but its internal
tree overlay also provides the hierarchy for the aggregation algorithm
\autocite{albrecht2004}. We use in the following a notation adapted from
Kademlia \autocite{maymounkov2002}.

\begin{longtable}[]{@{}rl@{}}
\toprule
\(k\) & maximum number of contacts per Kademlia segment
(\(k\)-bucket)\tabularnewline
\(x\) & a Kademlia leaf node ID (KID) of size \(B\)\tabularnewline
\(B\) & size of a KID in bits, e.g.~160\tabularnewline
\(x_i\) & KID of peer \(P_i\)\tabularnewline
\(d\) & node depth, i.e.~number of edges from the node to the tree
root\tabularnewline
\(\cSs(x,d)\) & subtree whose root is at depth \(d\) which contains leaf
node \(x\)\tabularnewline
\(\cS(x,d)\) & \emph{sibling subtree} whose root is the sibling node of
the root of \(\cSs(x,d)\)\tabularnewline
\bottomrule
\end{longtable}

\noindent
The leaf node identifiers \(x \in \{0,1\}^B\) (\(B\) bits) span the
Kademlia binary tree of height \(B\) and are denoted KID. Each peer
\(P_i\) joins the Kademlia overlay network using its KID defined as
\(x_i = \eta(t_i)\) with the authorization token \(t_i\) and the hashing
technique \(\eta\). This way, \(x_i\) depends on both \(P_i\)'s and
\(A\)'s key pair, so that \(x_i\) cannot be altered unilaterally
\autocite{baumgart2007}. \(B\) is chosen sufficiently large, so that
hash collisions leading to identical KIDs for distinct peers are very
unlikely. Consequently, the occupation of the binary tree is very
sparse.

Any node in the tree can be identified by its depth
\(d \in \{0,\ldots,B\}\) and any of its descendant leaf nodes with KID
\(x\). A \emph{subtree} \(\cSs(x,d)\) is identified by the depth \(d\)
of its root node and any of its leaf nodes \(x\). We overload the
subtree notation to designate as well the set of peers assigned to
leaves of the corresponding subtree. Further, we introduce \(\cS(x,d)\)
for the sibling subtree of \(\cSs(x,d)\), so that
\(\cSs(x,d) = \cSs(x,d+1)\cup\cS(x,d+1)\). The entire tree is
denoted~\(\cSs(x,0)\). We observe that
\(\forall d: P_i \in \cSs(x_i,d)\) and
\(\forall d: P_i \notin \cS(x_i,d)\).

In Kademlia, the distance \(d(x_i,x_j)\) between two KIDs is defined as
their bit-wise XOR interpreted as an integer. In general, a peer \(P_i\)
with KID \(x_i\) stores information on peers with \(x_j\) that are close
to \(x_i\), i.e.~for small \(d(x_i,x_j)\). For this purpose, \(P_i\)
disposes of a set denoted \(k\)-bucket of at most \(k\) peers
\(P_j \in \cS(x_i,d)\) for every \(\cS(x_i,d)\) with
\(d >0\).\footnote{Note that originally \autocite{maymounkov2002} the common prefix length $b$ is used to index $k$-buckets/sibling subtrees while we use the depth $d = b+1$ of the root of the subtree.}
See Fig.~\ref{fig:tree} for an example. The size of subtrees decreases
exponentially for growing depth \(d\). Consequently, the density of
known peers of corresponding \(k\)-buckets grows exponentially.

\tikzset{
  sarrow/.style    = {->, >=open triangle 90},
  darrow/.style    = {->, dashed,>=open triangle 90},
  point/.style     = {circle,draw,inner sep=0pt,minimum size=0pt},
  bigpoint/.style  = {fill=black,circle,draw,inner sep=0pt,minimum size=4pt},
  empty/.style     = {thick,gray,fill=black!25,circle,draw,minimum size=10pt},
  idleaf/.style    = {circle,draw,inner sep=1pt,minimum size=0.6cm,label={[yshift=0.8mm]below:$\cSs(x_i{,}3)$}},
  treeedge/.style  = {
    draw, 
    edge from parent path={(\tikzparentnode) -- (\tikzchildnode)}
  },
  mixp/.style      = {draw,rectangle,rounded corners}
  upbrace/.style   = {decorate,decoration={brace,mirror},thick},
  downbrace/.style = {decorate,decoration={brace},thick},
  highlightpath/.style = {very thick,draw=blue}
}

\begin{figure}
  \centering
  \begin{tikzpicture}
      [
          level 1/.style={sibling distance=40mm},
          level 2/.style={sibling distance=25mm},
          level 3/.style={sibling distance=13mm},
          edge from parent/.style={thick,draw=black!70},
          level distance=10mm,
      ]
      \node[point] (root) {}
      [
          every child node/.style={point}
      ]
          child {
              node (s01) {}
              child {
                  node (s02) {}
                  child {
                      node[empty] (s03)  {}
                      edge from parent
                      node[left] {0}
                  }
                  child[missing]
                  edge from parent
                  node[left] {0}
              }
              child {
                  node (s04) {}
                  child {
                      node[empty] (s05)  {}
                      edge from parent
                      node[left] {0}
                  }
                  child[missing]
                  edge from parent
                  node[right] {1}
              }
              edge from parent
              node[left=5pt] {0}
          }
          child {
              node {}
              child {
                  node {}
                  child {
                      node[idleaf] (x) {$x_i$}
                      edge from parent[highlightpath]
                      node[left] {0}
                  }
                  child {
                      node[empty] (s21) {}
                      edge from parent
                      node[right] {1}
                  }
                  edge from parent[highlightpath]
                  node[left] {0}
              }
              child {
                  node (s11) {}
                  child {
                      node[empty] (s12) {}
                      edge from parent
                      node[left] {0}
                  }
                  child {
                      node[empty] (s13) {}
                      edge from parent
                      node[right] {1}
                  }
                  edge from parent
                  node[right] {1}
              }
              edge from parent[highlightpath]
              node[right=5pt] {1}
          };
          \node[draw,dashed,fit=(s01) (s02) (s03) (s04) (s05),label=below:$\cS(x_i{,}1)$] (d3) {};
          \node[draw,dashed,fit=(s11) (s12) (s13),label=below:$\cS(x_i{,}2)$] (d2) {};
          \node[draw,dashed,fit=(s21),label=below:$\cS(x_i{,}3)$] (d1) {};
          \node[rectangle split, rectangle split parts=3, draw,rounded
          corners
          %,label=below:{buckets for peer $P_i$ with node ID $x_i$}
          ]
          at ($(x)+(0,-1.8)$)
          (bucket)
          {
              $k$-bucket for $d=3$
              \nodepart{second}
              $k$-bucket for $d=2$
              \nodepart{third}
              $k$-bucket for $d=1$
          };
          \draw[darrow] (d1.east) -- ++(0.4,0) |- (bucket.one east);
          \draw[darrow] (d2.east) -- ++(0.2,0) |- (bucket.second east);
          \draw[darrow] (d3.west) -- ++(-0.2,0) |- (bucket.third west);
  \end{tikzpicture}
  \caption{Example of Kademlia $k$-buckets for KID $x_i = 100$ assuming $B = 3$. The sparse tree is partitioned into subtrees $\cS(x_i,d)$ with root node at depth $d = 1,2,3$. The $k$-buckets for each $d$ contain at most $k$ peers $P_j \in \cS(x_i,d)$.}
  \label{fig:tree}
\end{figure}
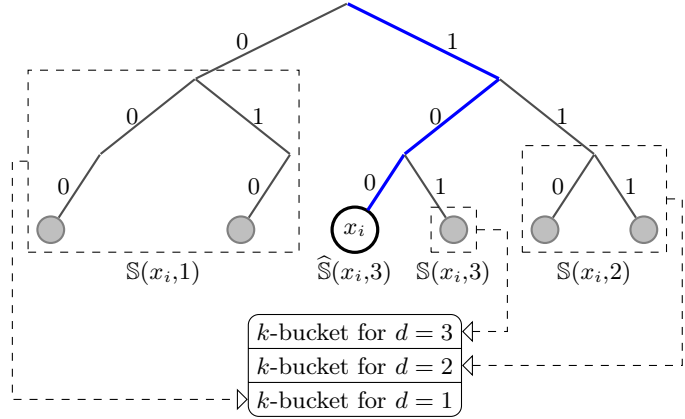

Kademlia ensures that the routing table, that is the set of all
\(k\)-buckets, is populated by peer lookup requests for random KIDs to
the closest already known peers. Requests are responded with a set of
closest, known peers from the routing table. One lookup might require
multiple, consecutive request-response cycles.

We assume that peers are either present or absent. Present peers join
the Kademlia overlay network within a given time interval and stay
responsive until their aggregation step is terminated. The aggregation
is carried out in \(B\) \emph{epochs}, one tree level at a time. Epochs
are loosely synchronized, because peers may have to wait for
intermediate aggregates to be computed in order to continue.

First, every peer \(P_i\) computes a container for its initial
aggregate. The container is assigned to represent the subtree
\(\cSs(x_i,B)\) with only \(P_i\).

In each epoch for \(d=B,\ldots,1\), every peer \(P_i\) requests from any
\(P_j \in \cS(x_i,d)\) the aggregate container of subtree
\(\cS(x_i,d)\). \(P_j\) responds with the demanded aggregate container.
With the received container of \(\cS(x_i,d)\) and the previously
obtained of \(\cSs(x_i,d)\), peer \(P_i\) computes the parent aggregate
using the aggregation operator \(\oplus\). Its corresponding container
is then assigned to the parent subtree \(\cSs(x_i,d-1)\). If
\(\cS(x,d) = \emptyset\) for any \(d\), the container of \(\cSs(x,d-1)\)
is computed only with the aggregate container of \(\cSs(x,d)\) from the
previous epoch.

After \(B\) consecutive epochs, peer \(P_i\) has computed the root
aggregate of the entire tree \(\cSs(x_i,0)\) that contains the initial
aggregates of all present peers. If all present peers are honest, the
root aggregate is complete and correct.

\section{Robust Aggregation}\label{sec:robust-aggregation}

\newcommand*\circled[1]{\tikz[baseline=(char.base)]{
            \node[shape=circle,draw,inner sep=1pt,font=\tiny] (char) {#1};}}
\newcommand*\circledtext[1]{\tikz[baseline=(char.base)]{
            \node[shape=circle,draw,inner sep=1pt,font=\scriptsize] (char) {#1};}}

The recursive aggregation introduced in
Section~\ref{sec:recursive-aggregation} is very vulnerable to aggregate
corruptions leading to erroneous root aggregates, and to illegitimate
requests compromising the confidentiality. Following the attack model
from \autocite{gambs2011}, we assume a minority of dishonest, Byzantine
peers entirely controlled by one adversary that aims to interrupt the
aggregation, manipulate root aggregates and increase its knowledge on
initial and intermediate aggregates. Byzantine peers can essentially
behave arbitrarily, but are assumed to be unable to prevent their
initial integration in the routing tables by honest peers.

To prevent Sybil attacks and arbitrary requests, all messages \(m\)
between peers are signed by the sender \(P_i\) using \(\sigma_i(m)\)
\autocite{baumgart2007}. For signature verification, the public key
\(pk_i\) and the token \(t_i\) must be either published (in the DHT) or
sent along with every signature. Henceforth, a peer \(P_i\) answers
aggregate requests for \(\cSs(x_i,d)\) only for peers
\(P_j \in \cSs(x_i,d)\) in the same subtree or \(P_j \in \cS(x_i,d)\) in
the sibling subtree. Consequently, peers cannot obtain more knowledge on
aggregates than strictly necessary to compute the root aggregate.

Further, peer signatures are employed to detect deviations from the
protocol. For every computed aggregate container of \(\cSs(x_i,d)\) with
hash \(h\) and counter \(c\), peer \(P_i\) produces an aggregate
container signature \(\sigma_i(h,d,c)\). A signature \(\sigma_i(h,d,c)\)
expresses the capacity of a peer \(P_i\) to compute the container
identified by its hash \(h\) and is consequently only valid for
containers of \(\cSs(x_i,d)\) for any \(d\).

In Fig.~\ref{fig:robust-pull}, we consider the steps of
\(P_j \in \cS(x_i,d)\) to produce for any \(P_i\) a \emph{confirmed
aggregate container} of \(\cS(x_i,d)\) backed by the signatures listed
below. Note that the necessary signatures depend on the subtree
configuration that can be explored by \(P_i\) using peer lookup
requests. Like for the recursive aggregation, \(P_j\) requests first the
sibling aggregate container~(\circledtext{1}) if
\(\cS(x_j,d+1) \neq \emptyset\). For \(|\cS(x_j,d+1)| < k\), the
corresponding \(k\)-bucket is exhaustive \autocite{maymounkov2002} and
the aggregate counter \(c\) must not exceed its size. \(k\)-buckets are
hardened against insertion of false contacts by requiring for all
\(P_q\) in lookup responses the proof of their KID \((pk_q,t_q)\). Then,
the so-called \emph{container candidate} for \(\cS(x_i,d)\) is
computed~(\circledtext{2}).

New is the \emph{confirmation} (\circledtext{3} and \circledtext{4}) to
acquire necessary signatures by otherwise redundant requests to peers in
the same subtree \(\cS(x_i,d)\). Candidates are exchanged solely among
peers of that subtree to allow for mutual confirmation.

\tikzset{
  xleaf/.style    = {circle,draw,inner sep=1pt,minimum size=0.6cm},
}

\begin{figure}
  \centering
  \begin{tikzpicture}
      [
          level 1/.style={sibling distance=40mm},
          level 2/.style={sibling distance=25mm},
          level 3/.style={sibling distance=13mm},
          edge from parent/.style={thick,draw=black!70},
          level distance=10mm,
          lfs/.style={font=\footnotesize},
      ]
      \node[point] (root) {}
      [
          every child node/.style={point}
      ]
          child {
              node (s01) {}
              child {
                  node (s02) {}
                  child {
                      node[xleaf] (s03)  {$x_i$}
                  }
                  child[missing]
              }
              child {
                  node (s04) {}
                  child {
                      node[empty] (s05)  {}
                  }
                  child[missing]
              }
          }
          child {
              node (sm1) {}
              child {
                  node (sm2) {}
                  child {
                      node[xleaf] (x) {$x_q$}
                      edge from parent
                      node[left] (smc2) {}
                  }
                  child {
                      node[xleaf] (s21) {$x_j$}
                  }
              }
              child {
                  node (s11) {}
                  child {
                      node[xleaf] (s12) {$x_l$}
                  }
                  child {
                      node[xleaf] (s13) {$x_p$}
                      edge from parent
                      node[right] (sm3) {}
                  }
              }
          };
          \node[draw,dashed,fit=(s01) (s03) (s04)] (bb) {};
          \node[below left, inner sep=2pt] at (bb.north west) {$\cSs(x_i{,}d)$};
          \node[draw,dashed,fit=(sm1) (x) (s13)] (b0) {};
          \node[below right, inner sep=2pt,align=left] at (b0.north east) {$\cS(x_i{,}d)=$\\$\cSs(x_j{,}d)$};
          \node[draw,dashed,fit=(s11) (s12) (s13)] (b1) {};
          \node[below right, inner sep=2pt] at (b1.north east) {$\cS(x_j{,}d+1)$};
          \node[draw,dashed,fit=(s13)] (b2) {};
          \node[below right, inner sep=2pt] at (b2.north east) {$\cS(x_l{,}d+2)$};
          \node[draw,dashed,fit=(x)] (b22) {};
          \node[below right, inner sep=2pt] at (b22.south west) {$\cS(x_j{,}d+2)$};
          % \node[draw,dashed,fit=(s11) (s12) (s13),label=below:$\cS(x_i{,}1)$] (d2) {};
          % \node[draw,dashed,fit=(s21),label=below:$\cS(x_i{,}2)$] (d1) {};
          % \path (sm1) edge[bend right=10,thick,->,>=stealth',shorten >=5pt, shorten <=20pt] node[label={[lfs,shift={(0,-5pt)}]above:pull}] {} (s01);
          \path (s11) edge[bend left=10,thick,->,>=stealth',shorten >=5pt, shorten <=10pt] node[label={[lfs,shift={(-9pt,4pt)}]below:\circled{1} pull}] {} (sm2);
          \path (sm2) edge[bend right=20,thick,->,>=stealth',shorten >=5pt, shorten <=5pt] node[label={[lfs,align=center,shift={(-6pt,2pt)}]right:\circled{2}\\compute}] {} (sm1);
          \path (s13) edge[bend right=30,thick,->,>=stealth',shorten >=5pt, shorten <=5pt] node[label={[lfs,align=right,shift={(-6pt,10pt)}]right:\circled{4}\\confirm}] {} (sm1);
          \path (x) edge[bend left=30,thick,->,>=stealth',shorten >=5pt, shorten <=5pt] node[label={[lfs,align=left,shift={(6pt,10pt)}]left:\circled{3}\\confirm}] {} (sm1);
  \end{tikzpicture}
  \caption{$P_j$ with $x_j$ produces a confirmed aggregate container of $\cS(x_i,b)$. This scheme applies to all tree levels with possibly large subtrees to request from.}
  \label{fig:robust-pull}
\end{figure}
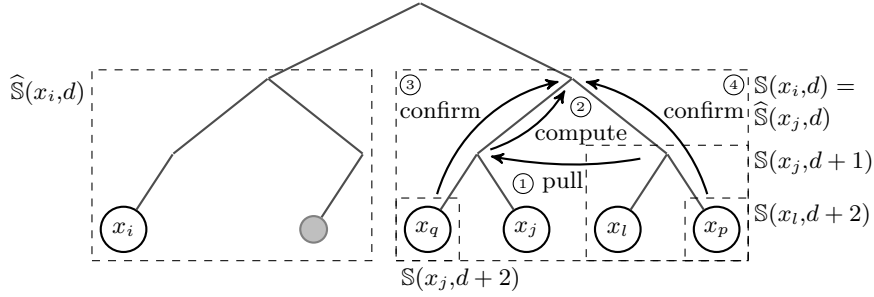

\noindent
\(P_i\) requires from \(P_j\) the following signatures with the
container for \(\cS(x_i,d)\):

\begin{enumerate}
\def\labelenumi{\arabic{enumi}.}
\tightlist
\item
  \(P_i\) requires the signature \(\sigma_j(h,d,c)\) on container hash
  and counter.
\item
  If \(c > 1\), there is at least one child aggregate with hash \(h_1\)
  and counter \(c_1\) and \(\sigma_j(h_1,d+1,c_1)\) must be provided.
\item
  If \(c > 1\) and \(c_1 > 1\), a confirmation request (\circledtext{3})
  is necessary to provide \(\sigma_q(h,d,c)\) from
  \(P_q \in \cS(x_j,d')\) with the smallest \(d' > d + 1\) for a
  non-empty subtree, ideally in the subtree \(\cS(x_j,d+2)\).
\item
  If \(c > 1\) and \(c_2 > 0\), \(P_j\) provides
  \(\sigma_l(c_2,d+1,h_2)\) acquired before (\circledtext{1}) as
  1.~signature.
\item
  If \(c > 1\) and \(c_2 > 0\), a confirmation request (\circledtext{4})
  is necessary to provide \(\sigma_l(h,d,c)\) if \(P_l\) for
  \(c_2 = 1\), and otherwise \(\sigma_p(h,d,c)\) from
  \(P_p \in \cS(x_l,d')\) with the smallest \(d' > d+1\) for a non-empty
  subtree, ideally in \(\cS(x_l,d+2)\).
\end{enumerate}

\noindent
The 1., 2.~and 4.~signature listed above are required already for
candidate containers and allow to detect dishonest peers during
confirmation. The 3.~and 5.~signature promote a consensus in
\(\cSs(x_j,d+1)\) respectively \(\cS(x_j,d+1)\).

The requests \circledtext{3} and \circledtext{4} provide additional
signatures, that may reveal dishonest peers deviating from the protocol.
Note that dishonest peers cannot influence which peers are requested to
avoid detection with certainty. For this, we focus on signatures
\(\sigma_e(h,d,c)\) and \(\sigma_e(h',d,c)\) of the same peer \(P_e\)
with equal counter \(c\) for distinct containers (\(h \neq h'\)) of the
same subtree. In case of \(c=1\), \(P_e\) derived from the protocol with
certainty, is as such detected as dishonest, and its signatures and
containers are discarded. A new candidate container without it is
computed. The same holds for \(c=2\), because \(P_e\) has not discarded
itself two distinct containers with \(c=1\) of the same peer, and alike
for \(c=3\). Without obvious proof for \(c>3\), we assume \(P_e\) to be
honest. The discarded signatures form a verifiable proof that is
attached to request responses for the newly computed (candidate)
container and stored in the DHT under the key \(\eta(x_e)\) if there was
none before. Detected dishonest peers are permanently removed from the
routing table.

With all required signatures, a candidate container of~\(\cS(x_i,d)\) is
confirmed and may be requested by peers in~\(\cSs(x_i,d)\). If the
candidate cannot be confirmed by a peer \(P_e\), a proof of former
deviation is looked up, and requests to other peers continue for a
limited number of tries. If \(P_j\) gathers this way a majority of
signatures for a different child container than those it has computed
earlier, \(P_j\) repeats the previous aggregation in order to correct or
confirm again its child container. If \(P_j\) gathers instead a majority
of signatures for a different child container than those it has
requested, \(P_j\) repeats the current aggregation in order to request
potentially a different sibling child container to use. Requests for
containers with \(c=1\) are not repeated to prevent revisions of initial
aggregates.

The administration step ensures that the global minority of dishonest
voters is randomly distributed over the tree. Hence, the implicit
majority vote on hashes is supposed to be decided by the local majority
of honest peers in the subtree. Note that a vote, and thus a honest
majority, is not required for subtrees with less than 4 peers, because
dishonest peers are detected and removed based on signatures on
containers.

If \(P_j\) can still not acquire all signatures, e.g.~due to a dishonest
peer \(P_e\) blocking the confirmation, \(P_j\) continues the
aggregation nevertheless and compensates the missing signature by both
child aggregate containers with all their signatures, so that the
aggregate computation of \(P_j\) can be reproduced. The confidentiality
of \(P_j\) and \(P_e\) is diminished to the same degree.

At last, the root aggregate container is confirmed by some additional
signatures to increase the confidence that it has been adopted by the
majority.

\section{Protocol Properties}\label{sec:properties}

Common security properties of online voting protocols
\autocites{ibrahim2003}{gambs2011} are considered using the attack model
of a dishonest minority from Section~\ref{sec:robust-aggregation}.

\paragraph{Eligibility}\label{eligibility}

The administrator is trusted to sign one authorization request for every
eligible peer. Without signature, peers cannot engage in the
aggregation.

\paragraph{Confidentiality}\label{confidentiality}

The protocol does not ensure secrecy of the initial aggregate due to the
necessity to share it at least once over a pseudonymous channel.
However, the access to the initial aggregate is limited to randomly
determined peers that acquire mostly partial knowledge, so that
confidentiality is ensured to a high degree. The pseudonymous channel
between peers augments further the confidentiality. The DHT is
ephemeral, distributes information evenly among peers, and vanishes when
peers disconnect after the aggregation. Potential data breaches are
therefore local and bounded in time.

\paragraph{Completeness and
Correctness}\label{completeness-and-correctness}

A local majority of dishonest peers in a subtree \(\cSs(x,d)\) with at
least 3 peers allow for manipulations of the corresponding aggregate
container. Manipulations of its counter \(c\) require further at least
\(k\) peers in \(\cSs(x,d)\). Hence, for a reasonably-sized global
dishonest minority, the protocol ensures that peers compute with high
probability root aggregates that are with high probability correct or
almost correct.

\paragraph{Verifiability}\label{verifiability}

Using requests, \(P_i\) can determine with high probability which root
aggregate has been confirmed by most peers and verify the chain of
container hashes to the hash of its initial aggregate.

\paragraph{Robustness and
Non-Interruptibility}\label{robustness-and-non-interruptibility}

The aggregation step is entirely distributed to equipotent peers. With
no weakest link, the influence of a reasonably-sized dishonest minority
is locally limited. The redundancy of the aggregate computation
increases exponentially in every epoch as aggregates become more
meaningful.

\vspace{1em}\noindent
The \emph{protocol complexity} is mostly inherited by the properties of
Kademlia, which have been studied \autocite{cai2013} and experimentally
confirmed as part of BitTorrent.

\paragraph{Message Complexity}\label{message-complexity}

For a network of \(n\) peers, a lookup requires with great probability
\landau{\log n} request-response cycles. Joining the network requires a
limited number of lookups and is thus as well of order \landau{\log n}.
With the consideration to estimate the number of empty \(k\)-buckets
from \autocite{cai2013}, the average number of container requests for
the basic aggregation is found to be \landau{\log n}.

\paragraph{Memory Complexity}\label{memory-complexity}

The memory required to store non-empty \(k\)-buckets is \landau{\log n}.
Further, the aggregation algorithm requires to store \landau{\log n}
received aggregate containers for non-empty sibling subtrees and perhaps
a limited number of alternatives in case of failing confirmations.
Hence, for a constant size of aggregate containers, the total memory
complexity is again \landau{\log n}.

\paragraph{Time Complexity}\label{time-complexity}

Intermediate aggregates for ancestor nodes are computed in sequence. For
a constant computation time per aggregate and with an upper limit to
request and confirm aggregates, the time complexity is \landau{\log n}.

\section{Experimental Confidentiality Analysis}\label{sec:simulation}

The protocol has been simulated on the basis of \texttt{kad}, an
implementation of
Kademlia\footnote{\url{http://kadtools.github.io/}, v1.6.2 released on November 29, 2016}
written in JavaScript with its extension \texttt{kad-spartacus}. For
each peer~\(P_i\), key pairs~\((pk_i,sk_i)\) are generated using
elliptic-curves cryptography. The KID \(x_i\) of each peer \(P_i\) is
derived by hashing \(pk_i\) first with SHA-256 and the result again with
RIPEMD-160. It is assumed that the use of \(pk_i\) instead of the
token~\(t_i\) leads to an equally random distribution of KIDs, so that
the administration step can be omitted in the simulation. A simulation
parameter allows to vary the generation of key pairs and consequently
the KIDs, so that different tree configurations can be tested.

After all \(n\) peers are instantiated, every \(P_i\) connects to the
Kademlia network using an initial contact \(P_{i-1}\). According to the
Kademlia protocol, peers update their routing table using lookup
requests. In our model, peers do not join or leave during the
aggregation, so that the routing table does not change hereafter. Once
all routing tables are complete, peers start the aggregation step like
detailed in Section~\ref{sec:recursive-aggregation}. The simulation does
not consider absent or dishonest peers.

If a peer receives a request for an intermediate aggregate that has not
yet been computed, the response is delayed. The aggregation steps in the
simulation use neither parallel requests nor timeouts for requests.

We consider the issue of confidentiality, and measure both the degree of
leakage of initial aggregates, and the concentration of knowledge on
initial aggregates. For that purpose, we assume that all initial
aggregates are distinct.

We define the \emph{leaked information} \(L_i\) of a peer \(P_i\) to be
the sum of the inverse of the counters of all containers that \(P_i\)
used to respond to aggregation requests. \(1/c\) denotes the probability
to correctly link the contained initial aggregate of~\(P_i\) to the
pseudonym of~\(P_i\), e.g.~an IP address. The leaked information \(L_i\)
is at least~1, because in a non-trivial aggregation with~\(n>1\),
\(P_i\) must respond at least once with its initial aggregate container
with~\(c=1\). In a perfectly balanced tree with~\(n = 2^B\) peers,
\(L_i\) is strictly smaller than~2: \[
  L_i = \sum^{B-1}_{n=0} \left(\frac{1}{2}\right)^n < 2
\]

Conversely, we define the \emph{received information} \(R_i\) of \(P_i\)
as the sum of \(1/c\) of all containers that \(P_i\) receives as
responses to its requests. In a perfectly balanced tree, \(R_i = L_i\).
We further introduce relative measures \(l_i = L_i/(n-1)\) and
\(r_i = R_i/(n-1)\) normed by the worst case that initial aggregates of
all other \(n-1\) peers are leaked/received. Fig.~\ref{fig:histograms}
shows the distribution of \(L_i\) and \(R_i\) for a simulation run with
\(n=1000\) peers. The simulation has been repeated with different tree
configurations without notable changes. In the examined case, the
relative leak to the network is \(l_i = \SI{0.24(9)}{\percent}\). The
relative received information \(r_i = \SI{0.24(10)}{\percent}\) is the
same with a slightly higher standard derivation.

The worst case is given by the least balanced tree configuration in
which \(|\cS(x_i,d)| = 1\) for all \(d \in \{B,\ldots,1\}\). That means
for the given~\(P_i\), every sibling subtree contains exactly one other
peer. Here, \(P_i\) learns in every epoch one initial aggregate with
certainty. However, such a tree allows for only \(B+1\) peers and every
additional peer decreases \(L_i\).

\pgfplotsset{%
  myhist/.style={%
    width=1\linewidth,
    height=0.7\linewidth,
    enlarge x limits=false,
    minor x tick num=3,
    ylabel={\# of peers},
    ymin=0,  
    % nodes near coords,
    ybar,
  },
  myhistaxis/.style={%  
    hist={%
        bins=20,
        data min=0,
        data max=10
    },
  },
}

\begin{figure}[!t]
  \begin{subfigure}[b]{0.49\textwidth}
    \begin{tikzpicture}%[trim axis left]
      \begin{axis}[myhist]
        \addplot +[myhistaxis] table [y index=0] {data/1000-01-leakedInformation.csv};
      \end{axis}
    \end{tikzpicture}
    \caption{Histogram of leaked information $L_i$.}
    \label{fig:histogram-leaked}
  \end{subfigure}
  ~
  \begin{subfigure}[b]{0.49\textwidth}
    \begin{tikzpicture}
      \begin{axis}[myhist]
        \addplot +[myhistaxis] table [y index=0] {data/1000-01-receivedInformation.csv};
      \end{axis}
    \end{tikzpicture}
    \caption{Histogram of received information $R_i$.}
    \label{fig:histogram-received}
  \end{subfigure}
  \caption{In a simulation with $n = 1000$, peers leak (\subref{fig:histogram-leaked}), respectively receive (\subref{fig:histogram-received}), information on initial aggregates depending on  the global distribution of peers on the binary Kademlia tree. $L_i$ peaks close to the theoretical value 2 of an optimally balanced tree. Only few peers leak significantly more. While the mean for $R_i$ is the same, the distribution is slightly different.}
  \label{fig:histograms}
\end{figure}
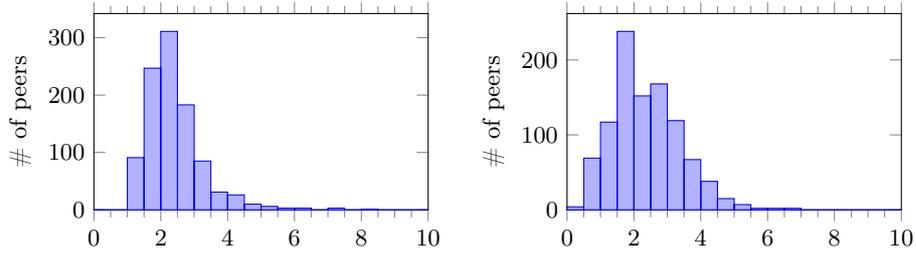

\pgfplotsset{%
  myhistboxaxis/.style={%  
    hist={%
        bins=25,
        data min=0,
        data max=25
    },
  },
}

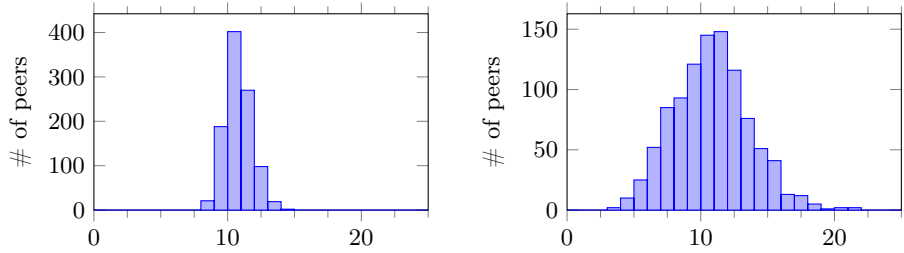
\begin{figure}[!t]
  \begin{subfigure}[b]{0.49\textwidth}
    \begin{tikzpicture}%[trim axis left]
      \begin{axis}[myhist]
        \addplot +[myhistboxaxis] table [y index=0] {data/1000-01-inbox.csv};
      \end{axis}
    \end{tikzpicture}
    \caption{Histogram of \# of received responses.}
    \label{fig:histogram-inbox}
  \end{subfigure}
  ~
  \begin{subfigure}[b]{0.49\textwidth}
    \begin{tikzpicture}
      \begin{axis}[myhist]
        \addplot +[myhistboxaxis] table [y index=0] {data/1000-01-outbox.csv};
      \end{axis}
    \end{tikzpicture}
    \caption{Histogram of \# of given responses.}
    \label{fig:histogram-outbox}
  \end{subfigure}
  \caption{In a simulation with $n = 1000$, the number of given (\subref{fig:histogram-outbox}) and received (\subref{fig:histogram-inbox}) responses has been recorded for every peer. While the distribution of received responses is very sharp, the distribution for given responses is twice as broad. In the Kademlia routing tables, some peers are more often represented than others.}
  \label{fig:histograms-box}
\end{figure}

Moreover, the load on peers measured by the number of received and given
responses has been examined. The histograms in
Fig.~\ref{fig:histograms-box} indicates that no peer receives
significantly more load than others---a property that has been shown for
Kademlia before.

Eventually, the average number of requests per peer simulated with
different numbers of peers \(n\) up to \(n = 1000\) confirmed the
theoretical message complexity of \landau{\log n} shown in
Section~\ref{sec:properties}.

\section{Conclusion}\label{sec:conclusion}

We considered the fundamental problem of large-scale confidential
aggregation, and proposed the distributed aggregation protocol
\textsc{Advokat}. It prioritizes system wide properties like scalability
and robustness over perfect completeness, correctness or full secrecy of
initial aggregates.

The aggregation step is distributed to entirely equipotent peers which
improves the robustness in face of all sorts of attacks and reduces the
reliance on institutional trust. Peers may choose their trusted protocol
implementation. Cryptography is only employed to manage authorization
and ensure integrity, but not to ensure secrecy, which renders the
protocol easier to understand and independent of hardness-assumptions
common in cryptography. Due to the even distribution of data and the
ephemeral nature of the network, the risk of global or targeted leaks
after the aggregation is eliminated. With its global message complexity
of \(\landau{n\log{n}}\), it outperforms SPP
with~\landau{n\log{n}^3}~\autocite{gambs2011} and DPol
with~\landau{n\sqrt{n}}~\autocite{guerraoui2012} which both provide
instead stronger confidentiality.

We showed that the protocol offers a high level of confidentiality
though at least comparable to postal voting with trusted authorities.
For large \(n\), it is very unlikely that the initial aggregate of a
given peer is revealed, which might be acceptable for many applications.
Completeness and correctness can be compared to paper-based voting. It
is possible that few initial aggregates are manipulated or not counted,
but not at a global scale and not often. An individual verification
allows to detect manipulations.

The universal protocol algebra supports a wide range of applications,
e.g.~distributed lottery, aggregation of sensible healthcare data, or
all sorts of reduce operations. Turning our protocol into a solution
that can be adopted in practice will require some effort. Foremost, a
formal definition of completeness and correctness must be introduced so
that upper limits of their manipulations depending on the ratio of
dishonest peers in the attack model from
Section~\ref{sec:robust-aggregation} can be formulated. Further, the
influence of churn of Byzantine peers on the routing tables must be
analysed and, if necessary, restricted to allow for Byzantine peers with
no assumptions.

\subsubsection{Acknowledgments}\label{acknowledgments}

The authors thank Stéphane Frénot, Damien Reimert, Aurélien Faravelon,
Pascal Lafourcade and Matthieu Giraud for fruitful discussions on
distributed voting protocols and attack vectors.

% \printbibliography
\section*{\bibname}
\printbibliography[heading=none]

\vfill
\end{document}